\documentstyle[multicol,aps,prl]{revtex}

\begin{document}
\widetext
\draft

\title{ High Rayleigh number turbulent convection in a gas near the gas-liquid 
                            critical point. } 
\author{ Shay Ashkenazi and Victor Steinberg}

\address{Department of Physics of Complex Systems\\
         The Weizmann Institute of Science, 76100 Rehovot, Israel}

\date{\today}
\maketitle

\begin{abstract}
 $SF_6$ in the vicinity of its critical point was used to study turbulent 
convection up to exceptionally high Rayleigh numbers, $Ra$, (up to 
$5\cdot 10^{14}$) and 
to verify for the first time the generalized scaling laws for the heat
 transport and the large scale circulation velocity as a function of $Ra$ 
 and the Prandtl 
 number, $Pr$, in very wide range of these parameters. The both scaling laws
 obtained are consistent with theoretical predictions by B.Shraiman 
 and E.Siggia, Phys. Rev. {\bf A 42}, 3650 (1990).

\end{abstract}

\begin{multicols}{2}
\narrowtext
Turbulent convection has recently attracted a lot of attention due to 
possibility to tune
 and control the relevant parameters with unbeatable precision. The fundamental
 aspect of the turbulent convection is a competition between buoyancy and shear
 \cite{sig}. 
An important issue in our understanding of the convective turbulence is 
experimental test of theoretical predictions on the Prandtl number, $Pr$, 
dependence of global characteristics, such as nondimensional heat transport, 
$Nu(Ra,Pr)$, and Reynolds number
of the large scale circulation flow, $Re(Ra,Pr)$. The difficulty to measure the 
$Pr$ dependence in
turbulent convection arises from the fact that in conventional fluids it cannot
 be varied substantially but only by changing the fluid which is not always a 
 simple task.
 The verification of these scaling relations will be 
a crucial test of the theory.\\
  In this Letter we present results on the Pr dependence of global properties 
 of the flow, namely, global heat transport, characterized by $Nu$, and large 
 scale circulation velocity, characterized by $Re$.\\
There are two possibilities to reach high $Ra$ number convection regime: 
either 
to increase temperature difference across the cell or to vary physical 
parameters, which appear in the expression for $Ra=\frac 
{g\alpha \Delta T L^3}{\kappa\nu}$, where $\Delta T$ is the temperature 
difference across the cell, $g$ is the gravitational constant,
$\alpha$ is the thermal expansion, and $\kappa$ and $\nu$ are the thermal 
diffusivity and kinematic viscosity, respectively. The latter possibility was 
used successfully in compressed gases ($He$ and $SF_6$) at almost 
constant value of $Pr$ about one\cite{hesl,cast,wu,til}.
 Another system in which this 
method of variation 
of $Ra$ number in a wide range can be used, is a gas near its gas-liquid
critical point(CP). It was realized long time ago that heat transport is 
enhanced dramatically near $T_c$ \cite{stein}. But it was only recently
shown experimentally that the critical temperature difference for the 
convection onset
decreases drastically with the distance from $T_c$ due to strong variations
 of thermodynamical and kinetic properties of a gas in the vicinity of 
 CP \cite{mich}. Singular
 behaviour of the thermodynamic and kinetic properties
of the fluid near $T_c$ provides the opportunity both to reach extremal
 values of the control parameter $Ra$ and to scan $Pr$
 over an extremely wide range . All these 
features make the system unique in  this respect\cite{shay,pap}.\\
However, the most exciting aspect of the system is the possibility to perform 
Laser Doppler velocimetry(LDV) measurements that we recently demonstrated 
\cite{shay,pap}. 
Small temperature differences used to reach high $Ra$, lead to rather small 
fluctuations
of refraction index in the flow. It gives us the possibility to use a standard
LDV technique. On the other hand, the problem of seeding particles in a 
close flow of a gas, which coagulate and sediment rather fast, turns out 
to be unsolvable
 at the moment. Fortunately, we discovered that the existence of the critical 
 density fluctuations provided us the possibility to perform LDV measurements 
 of the velocity field in a rather wide range of closeness to the CP 
 between $3\cdot 10^{-4}$ and $10^{-2}$ in reduced temperature $\tau=
 (\overline T-T_c)/T_c$, where $\overline T$ is the mean cell temperature.
  The upper limit is defined by a small 
  scattering amplitude, and the lower limit- by multiscattering in the large 
  convection cell \cite{shay,pap}.\\
At the same time those obvious advantages come together with limitations and 
new features which should be taken into account and studied. Strong dependence
 of gas properties on the closeness to $T_c$ manifests itself in 
 nonuniform distribution of density in a gravitational field and variation of 
 coefficients in the Navier-Stokes equation with temperature and density. The 
 former, the well-known gravity effect\cite{mold}, can be significant 
 even at $\tau \approx 10^{-3}$ and 10 cm height cell: the density difference
 across the cell reaches 1\% that leads to rather significant variations in 
 the fluid properties. Fortunately, a temperature gradient compensates the 
 gravity
 when heating from below\cite{stein}, and strong convection reduces the
  characteristic size of
nonuniformity to one of about the boundary layer height. This leads to reduction
 of the density 
 nonuniformity to a tolerant level much below 0.1\%.\\
  In the Boussinesq
 approximation fluid properties are assumed to be constant despite the 
 temperature gradient across the cell, except for the buoyancy  term.
  According to \cite{wu1} the degree of deviation from the Boussinesq
   approximation is adequately described by the ratio of the temperature drop
across the top boundary layer to the temperature drop across the bottom 
boundary layer. As the experiment shows \cite{wu1}, significant deviations in 
the measurements of global transport properties of the turbulent convection
occurs if this ratio reduces below 0.5. According to our estimates measurements
at $Ra$ up to $10^{15}$ can be done while non-Boussinesq effects are still 
relatively small. In the data presented the temperature drop ratio was above
0.7. So small deviations from the scaling behaviour were observed at highest 
$Pr$ and $Ra$.\\
 There are another
aspects, related to the proximity to $T_c$ such as compressibility (besides
 adiabatic gradient) and breaking 
 down of hydrodynamic description due to macroscopical size of thermodynamical 
 fluctuations. Simple estimates show that both these factors do not play any 
 significant role in the range of parameters under studies \cite{shay}.\\ 
 Contrary to the compressed gas convection\cite{cast,wu,til}, here one
  cannot cover wide range of $Ra$ for
 one value of $Pr$. On the other hand, the data presented cover the range of 
 $Ra$ from $10^{10}$ at low values of $Pr$ ($Pr=8$) far from $T_c$ and 
$\Delta T=12 mK$ up to the largest attainable in a laboratory $Ra=5\cdot 
10^{14}$ at high values of $Pr$ close to $T_c$. The whole range of the 
reduced mean temperature was $4\cdot 10^{-2} > \tau > 2\cdot 10^{-4}$.
And finally, as a result of large compressibility and relatively large cell 
height the adiabatic temperature gradient was observed\cite{ll}. This effect 
will be discussed below.\\
The experiment we present here, was done with a high purity gas 
$SF_6$ (99.998\%) in the vicinity of $T_c$ and at the 
critical density($\rho_c=730 kg/m^3$). This fluid was chosen due to the 
relatively low critical temperature
 ($T_c=318.73$ K) and  pressure($P_c=37.7$ bar) and well-known 
thermodynamic and kinetic properties far away and in the vicinity of 
CP. This gas was widely used to study the equilibrium critical 
phenomena. The cell is a box of a cross-section
76x76 $mm^2$ formed by 4 mm plexiglass walls, which are sandwiched 
between a Ni-plated 
mirror-polished copper bottom plate and a 19 mm thick sapphire top plate of 
$L=105$ mm apart. \\ 
The cell is placed inside the pressure vessel with two side thick plastic 
windows to withstand the pressure difference up to 100 bar. So the cell had 
 optical accesses from above through the sapphire window and from the sides. 
 They were used for both shadowgraph flow visualization  
  and for LDV. The pressure vessel was placed
 inside a water bath which  was stabilized with rms of temperature fluctuations
  at the level of 0.4 $mK$. The gas pressure was continuously measured 
 with 1 $mbar$ resolution by the absolute pressure gauge. Together with
 calibrated 100$\Omega$ platinum resistor thermometer they provide us the
 thermodynamic scale to define the critical parameters of the fluid and then to
 use a parametric equation of state developed recently for $SF_6$\cite{seng}.\\
 We should point out here that it is not obvious at all that the critical 
 parameters, defined at equilibrium conditions, can be used for a strongly 
 non-equilibrium state to define closeness to  CP. Contrary, as suggested by 
 theory\cite{onu}  and experiment\cite{pine}, turbulent flow in a binary  
 mixture near its consolute CP can suppress the critical concentration 
 fluctuations. At 
 $Re$ comparable with that achieved in our experiment, Pine et al\cite{pine}
 observed the critical temperature depression up to 50 $mK$. We used the  
 enhancement of the heat transport(or decrease in the temperature difference 
 across the cell at the fixed heat flux) near  CP as an indicator of the
  closeness to  CP. The critical pressure and temperature obtained by this 
  procedure agree within $\pm 10 mK$ with those defined independently by light 
  scattering at equilibrium conditions\cite{shay}.\\  
 The heat transport measurements were conducted at the fixed heat flux and 
 temperature of the top sapphire plate, while the bottom temperature was 
 measured by a calibrated thermistor epoxied in it. Local 
temperature 
 measurements in a gas were made by three $125\mu m$ thermistors suspended on 
glass fibers in the  interior of the cell (one at the center, and two about 
half way from the wall). Local vertical component velocity measurements at 
about $L/4$ from the bottom plate were conducted by 
using LDV on the critical density fluctuations\cite{shay,pap}. Shadowgraph 
visualization was used mostly to get qualitative information about  
structures and characteristic time and length scales in the flow mostly of the 
top and bottom boundary layers.\\
 From the measurements of the heat transport and, particularly, of the 
 velocity we realized that at a small but finite  temperature differences, much 
 larger than defined by the critical $Ra$ for convection onset,
 there exists a mechanically stable state. This temperature difference
  $\Delta T_{ad}$ is defined by the adiabatic temperature gradient for the 
  convection onset.
 The latter results from a fluid compressibility and is defined as
$ \Delta T_{ad}/L=gT\alpha{C_p}^{-1}$\cite{ll}, where $C_p$ is the heat 
capacity at constant pressure.
 As follows from simple thermodynamics and critical 
 divergences of thermodynamical parameters  as $T$ approaches $T_c$,
  the adiabatic temperature gradient saturates at 
a finite value $\Delta T_{ad}/L=g\rho (\partial P/\partial T)_v^{-1}$, which for
 $SF_6$ and $L=10.5$ cm gives $\Delta T_{ad}=9.5 mK$ for $\tau < 10^{-2}$, 
  So with available 
 temperature stability and resolution it can be  measured. Then, as shown in 
 \cite{stein}, $Ra$ should be modified as $Ra=\frac{gL^3\alpha 
 (\Delta T-\Delta T_{ad})}{\nu \kappa}$.\\
 The most sensitive probe to detect the convection onset in our case was the 
 local velocity measurements. The results of the mean vertical velocity
  $V_m$ and rms of vertical 
 velocity fluctuations $V_s$ measurements at $\tau=8\cdot 10^{-4}$ are 
 presented in Fig.1. Both signals show clear transition at 
 $\Delta T_{ad}=9.5 \pm 0.5 mK$, that agrees well with the theoretical value
 \cite{shay,cast1}.
 These measurements were done for several values of the reduced temperature in 
 the range $8\cdot 10^{-3}> \tau >2\cdot 10^{-4}$. It was found that 
 $\Delta T_{ad}(\tau)$ is constant and independent of $\tau$ in agreement with 
 the theory\cite{stein}. This value of $\Delta T_{ad}$ was used further 
 to correct the $Ra$ values. As a result of this correction for each $Pr$ both 
 $Nu$ vs $Ra$(Fig.2) and $V_s$ vs $Ra$(Fig.3) dependences show scaling laws in 
 rather wide range of $Ra$\cite{shay}.\\
  The heat transport measurements were
 done in the range $1.6\cdot 10^{-2}> \tau > 2\cdot 10^{-4}$. In order to keep
 $Pr$ constant, $\tau$ was kept constant, while $\Delta T$ changed during 
  the measurements. 
 The range of $\tau$ was limited from higher temperature end by mechanical 
 stability of 
 constructing materials (mostly plexiglass) and from lower end by the 
 temperature stabilization of the system. Heat transport measurements were also
 conducted
 far away from CP at $P=20 bar$, $\overline T=303 K$ and the density 
 $\rho=0.18 g/cm^3$, that corresponds to $Pr=0.9$, and at $P=50 bar$,
  $\overline T=323 K$ and $\rho=1.07 g/cm^3$, that corresponds to $Pr=1.5$. 
  The data far away from CP cover the range of $Ra$ between $10^9$ and 
  $5\cdot 10^{12}$ with the temperature differences across the cell from about
  $0.1 K$ till $10 K$. By making all appropriate corrections for heat losses 
  through the lateral walls, gas and insulation outside the cell, and for a
  temperature drop across the top sapphire plate, we found that the data on
  the heat transport far away from CP are in a good agreement with a 2/7 law 
for $ 10^9\leq Ra \leq 5\cdot 10^{12}$\cite{cast}. These data combined with the
 heat transport measurements in the same cell in the vicinity of CP fot higher
 values of $Pr$ can be scaled and presented in a power law form: 
 $Nu=0.22 Ra^{0.3 \pm 0.03} Pr^{-0.2 \pm 0.04}$ (Fig.4). Just a few points at 
 the highest $Pr$ and $Ra$ deviate from this scaling due to the non-Boussinesq 
effect. This scaling is consistent with predictions of ref\cite{sig}.\\
 We visualized a structure of the large scale flow by using shadowgraph 
 technique through the top and side windows. Large diameter with narrow focal 
 width imaging optics enabled us to visualize narrow slices across the cell 
 and scan it in both vertical and horizontal directions. The flow was recorded
  at various $Ra$ and $Pr$. This visualization confirmed the picture of 
  up- and down-going circulation jet flow along the cell diagonal,
 which forms the top and bottom  turbulent boundary layers. The bulk of the cell
  appears homogeneous. The turbulent character
 of the flow at the boundary layers is seen as a rapid and "violent" horizontal
 motions \cite{shay,pap}.\\
  The circulation frequency which corresponds to the travel
 time of a fluid element passing through one cycle of the large eddy, is 
 observed as a peak in the power spectra for both the velocity and temperature 
 fluctuations(see the insert in Fig.5). We also measured directly by LDV 
 the large scale circulation velocity.  
 However, the best results were obtained by extracting the peak frequency from
 the velocity power spectra at various $Ra$ and $Pr$ 
 which can be presented in 
 a power law form: $Re=2.6 Ra^{0.43 \pm 0.02}Pr^{-0.75 \pm 0.02}$ (Fig.5). 
 Here $Re=4 f_pL^2/\nu$, and $f_p$ is the peak frequency. The data from the low 
 frequency peak in the temperature spectra far away and close to CP and from 
 LDV measurements are consistent with this scaling law. We would like to 
 point out that $Re$ reaches values up to $10^5$ at the highest values of 
 $Ra$.\\
 The dependence of $Re$ on $Pr$ has almost the same exponent (within the 
 measurement uncertainty) as one found recently in convection in helium for
  much 
 narrow range of $Pr$\cite{chav}, and agrees rather well with the theoretical
 prediction 5/7 in ref\cite{sig}. However, the scaling of $Re$ with $Ra$ differs
 significantly and close to 3/7 rather to 0.5\cite{chav}.\\
 We can also verify consistency of the exponents of the both scaling laws
  obtained by using one
 of them, e.g. for $Re$, and the exact relation for the dissipation in a bulk 
 turbulent regime $PrRa(Nu-1)\sim Re^3Pr^3$\cite{sig}. The resulting scaling 
 relation $Nu \sim Ra^{0.29}Pr^{-0.25}$ is consistent with that found 
 experimentally. Together with the unified scaling law in the whole range of 
 $Ra$ and $Pr$ under studies for both transport mechanisms it suggests that
  a single mechanism is responsible for these scaling laws. Moreover, our data
  did not indicate any signature of the transition to the asymptotic regime in
  the heat transport, discovered recently at relatively low $Pr$ and 
  $Ra>10^{11}$. There are possible two explanations to this fact: either
   at higher $Pr$ the transition occurs at higher $Ra$, or the 
   scatter in the data for different $Pr$ does not allow us to observe
   the transition.\\
 In conclusion, we present for the first time the generalized scaling laws 
  of the heat transport and the large scale flow in respect to $Ra$ and $Pr$ in
  a wide range of variation of there parameters. Both scaling laws are 
  consistent with the theoretical predictions of \cite{sig}. We believe that 
  in the light of the recent theoretical reconsiderations of these scaling 
  laws\cite{lohse} these data provide new insight on the turbulent 
  convection.\\  
  This work was partially supported by the Minerva Foundation and the Minerva 
 Center for Nonlinear Physics of Complex Systems. VS is greatful for support 
 of the Alexander von Humboldt Foundation. \\

\begin{figure}
\caption{LDV measurements of the mean $V_m$(squares) and rms 
$V_s$( circles) vertical velocity component.}

\label{figa}
\end{figure}
\begin{figure}
\caption { $Nu$ vs $Ra$: open squares-uncorrected, and full squares-corrected
 data at $Pr=27$.}

\label{figb}
\end{figure}
\begin{figure}
\caption{ $V_s$ vs $Ra$: circles-uncorrected, and triangles-
 corrected data at $Pr=93$. } 

\label{figc}
\end{figure}

\begin{figure}
\caption{ $Nu$ vs $Ra$ for different $Pr$. }

\label{figd}
\end{figure}
\begin{figure}
\caption{ $Re$ of the large scale circulation vs $Ra$ for different $Pr$:
 circles- 27; up-triangles- 45; down-triangles- 93,  
 diamonds- 190. The insert: power spectra density of vertical velocity 
 fluctuations at $Ra=3\cdot 10^{12}$ at $Pr=93$.}

\label{fige}
\end{figure}

\end{multicols}


\begin{references}
\bibitem{sig} B. I. Shraiman and E. D. Siggia, {\it Phys. Rev.} {\bf A 42}, 
 3650 (1990); E. D. Siggia, {\it Annu. Rev. Fluid Mech.} {\bf 26}, 137 (1994).
\bibitem{hesl} F. Heslot, B. Castaing, and A. Libchaber, {\it Phys. Rev.} 
            {\bf A 36}, 5870 (1987).
\bibitem{cast} B. Castaing et al., {\it J. Fluid. Mech.} {\bf 204}, 1 (1989).

\bibitem{wu} X. Wu, L. Kadanoff, A. Libchaber, and M. Sano, {\it Phys. Rev. 
        Lett.} {\bf 64}, 2140 (1990).

\bibitem{til} A. Tilgner, A. Belmonte, and A. Libchaber, {\it Phys. Rev} 
{\bf E 47}, R2253 (1993).
\bibitem{stein} M. Gitterman and V. Steinberg,  {\it High Temperature (USSR)} 
  {\bf 8}, 754 (1970); {\it J. Appl. Math. Mech.(USSR)} {\bf 34}, 305 (1971).
\bibitem{mich} M. Assenheimer and V. Steinberg, {\it Phys. Rev. Lett.}
 {\bf 70}, 3888  (1993).
\bibitem{shay} Sh. Ashkenazi, Ph.D. thesis, Weizmann Institute of Science,
      Rehovot, Israel, 1997.
\bibitem{pap} Sh. Ashkenazi and V. Steinberg, to be published.
\bibitem{mold}M. R. Moldover, J. V. Sengers, R. W. Gammon, and R. J. Hocken,
{\it Rev. Mod. Phys.} {\bf 51}, 79 (1979).
\bibitem{wu1} X. Z. Wu and A. Libchaber,{\it Phys. Rev.} {\bf A43}, 2833 (1991).
\bibitem{ll} L. D. Landau and E. M. Lifshitz, {\it Fluid Mechanics}, Pergamon 
 Press, Oxford, 1984.

\bibitem{seng} A. Abbaci and J. Sengers, {\it Technical Report, BN1111}, 
Institute for Physical Sciences and Technology, University of Maryland,
 USA,1990.
\bibitem{onu}R. Ruiz and D. R. Nelson, {\it Phys. Rev.} {\bf A 23}, 3224 
(1981);  A. Onuki {\it Phys. Lett.} {\bf 101A}, 286 (1984).
\bibitem{pine} D. J. Pine, N. Eswar, J. V. Maher, and W. I. Goldburg, 
 {\it Phys. Rev.} {\bf A 29}, 308 (1984).

\bibitem{cast1} It was the first direct laboratory measurements of 
the adiabatic temperature gradient (see ref\cite{shay}). About at the same 
time and independently it was also observed in low temperature convection
in helium
( X. Chavanne et al. {\it J. Low Temp. Phys.} {\bf 104}, 109 (1996) and 
B. Castaing, private communication (1996)). Recently more precise data on the 
convection onset in a compressible fluid- $He^3$ near CP -were presented 
by A. Kogan, D. Murphy, and H. Meyer (preprint,1999).
\bibitem{chav} X. Chavanne et al. {\it Phys. Rev. Lett.} {\bf 79}, 3648 (1997).
\bibitem{lohse} S. Grossmann and D. Lohse, preprint,1998.
\end{references}
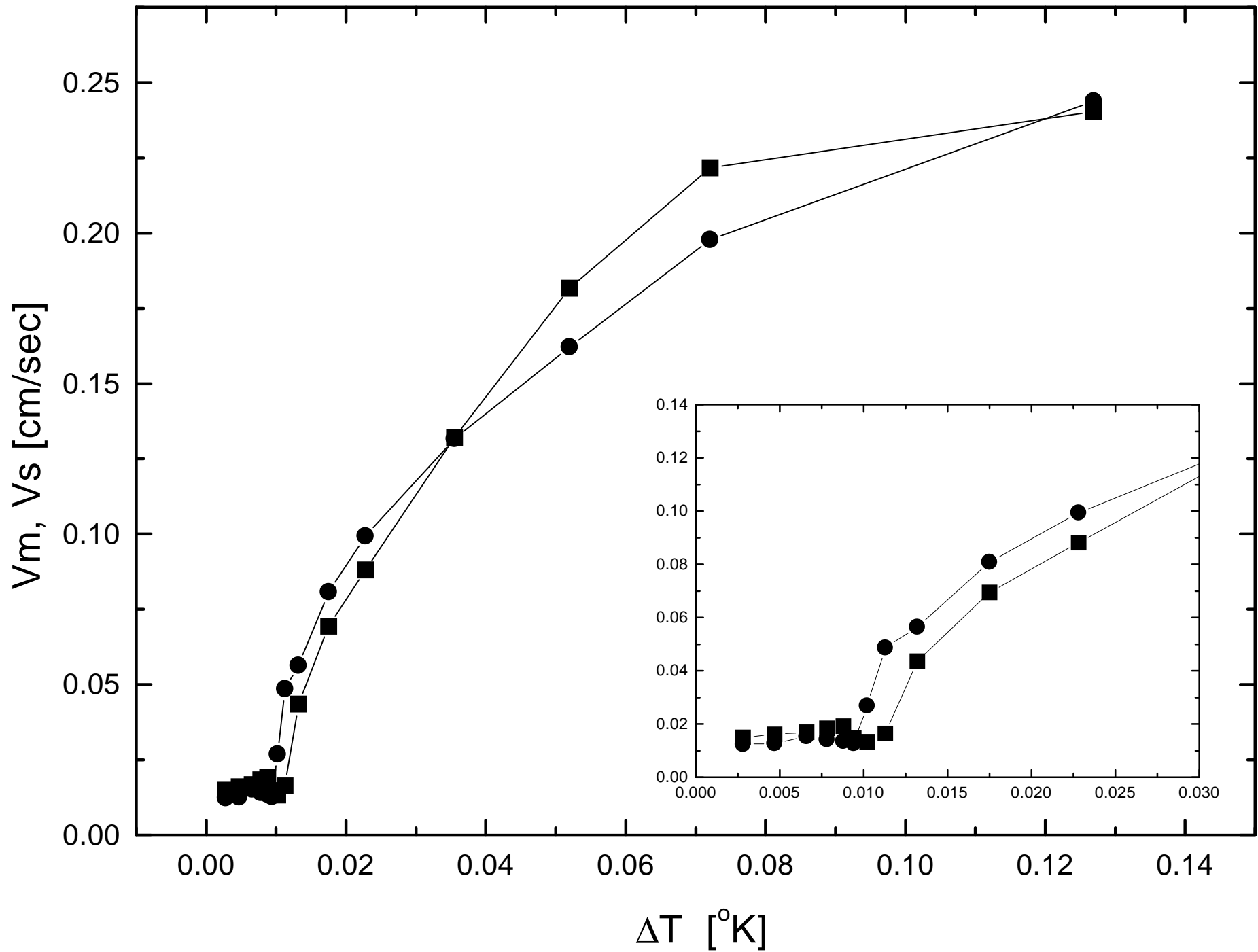
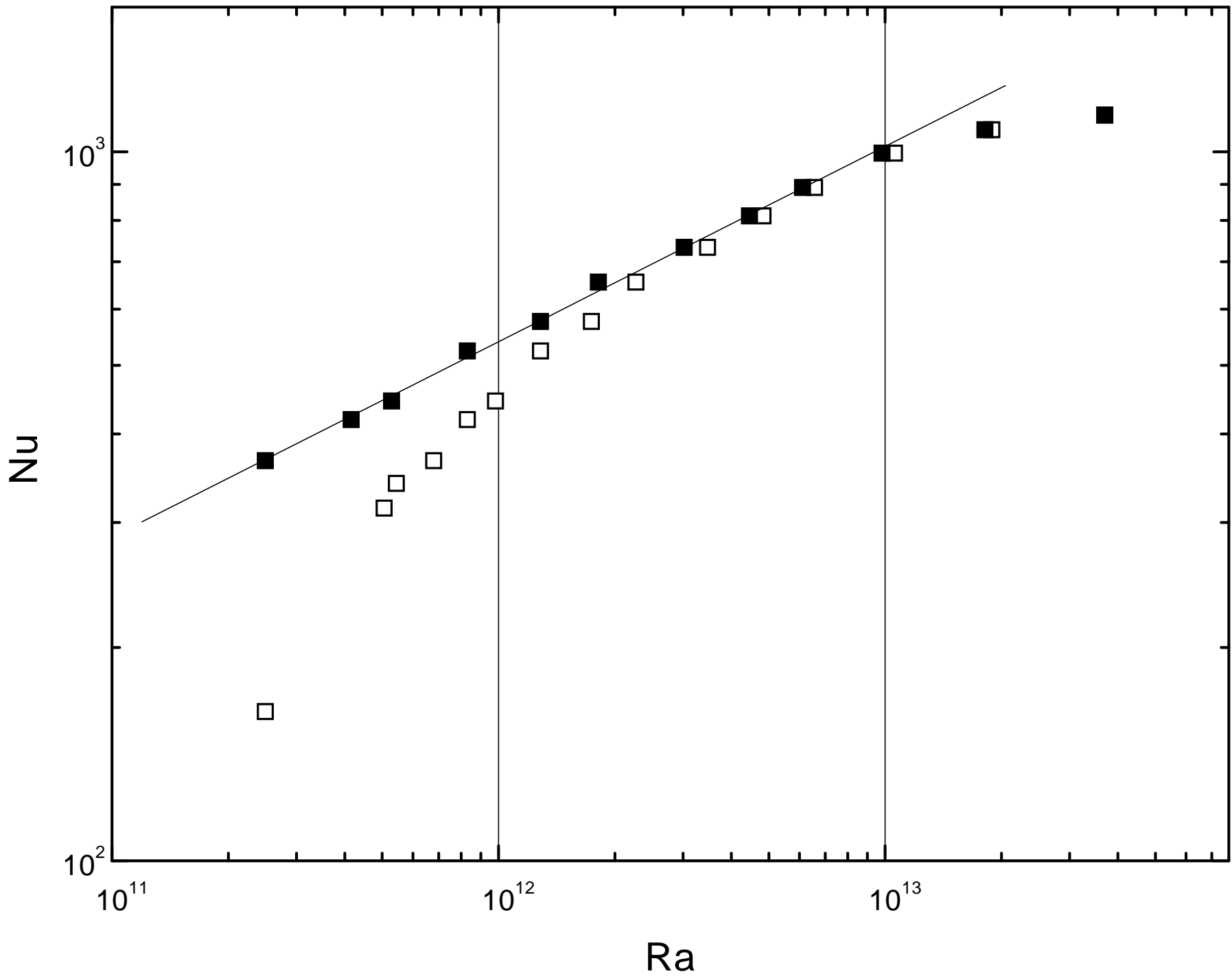
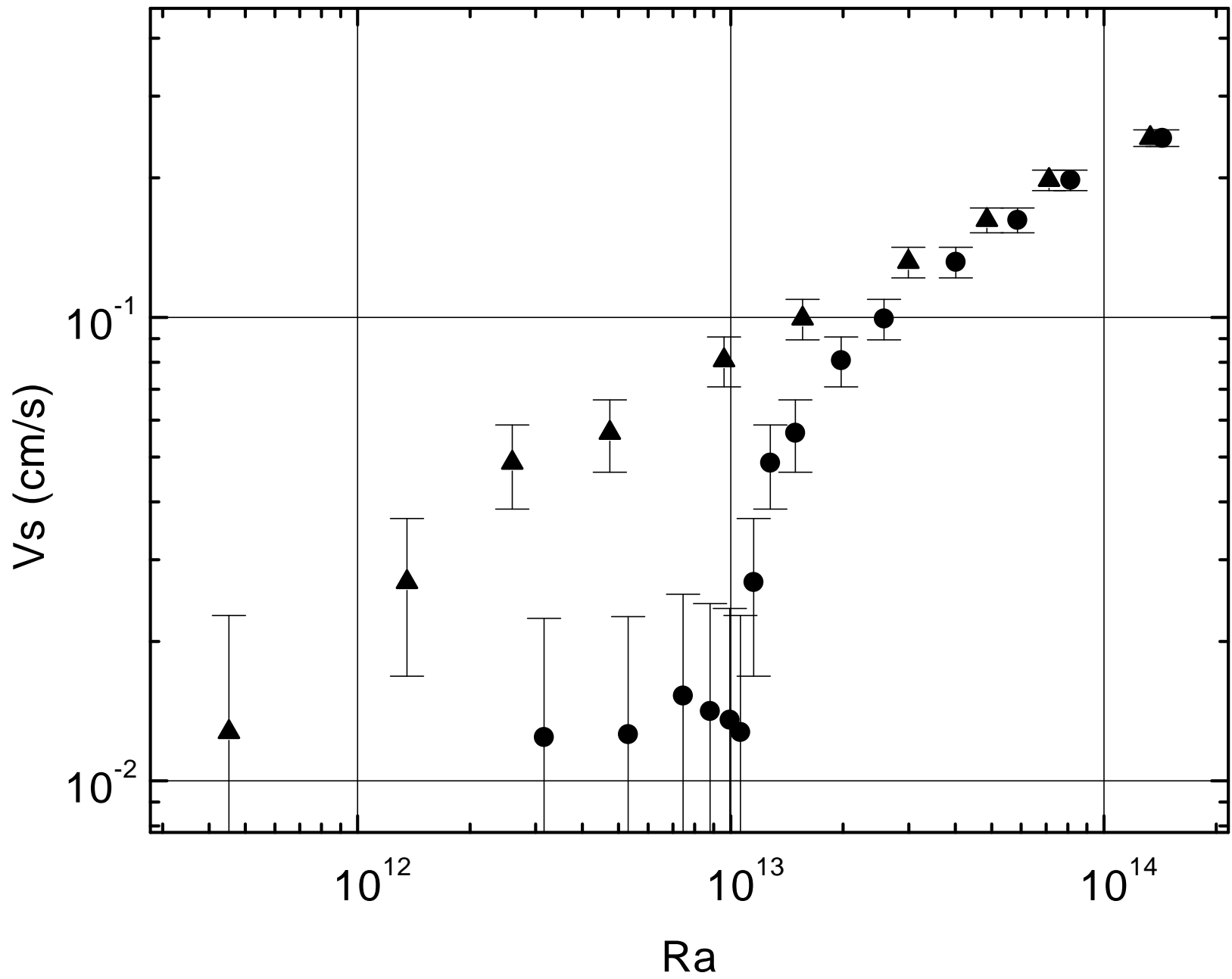
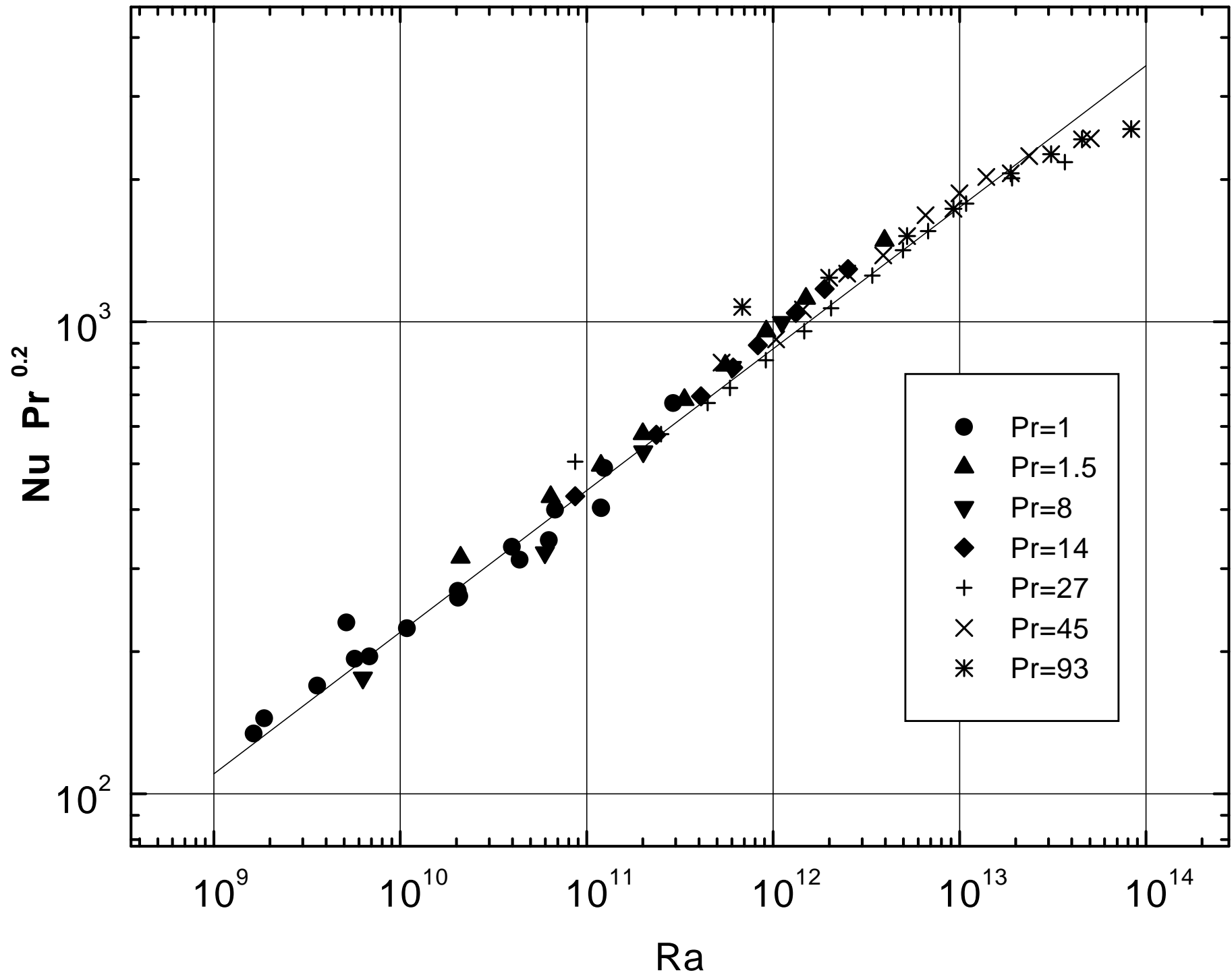
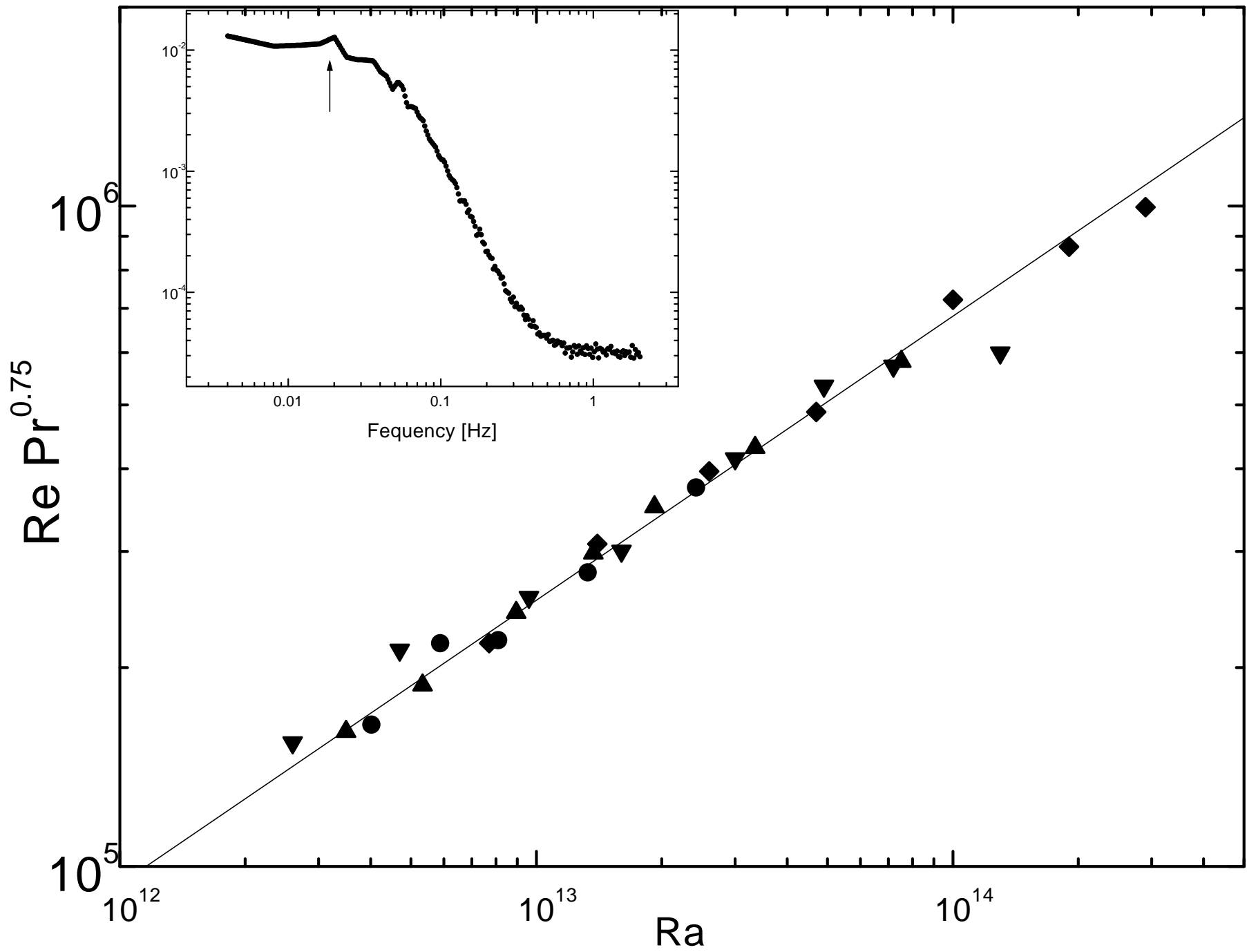
\end{document}